# Monitoring LMXBs with the Faulkes Telescope

**Fraser Lewis**
*Faulkes Telescope Project*
*School of Physics and Astronomy, Cardiff University, 5 The Parade, Cardiff, Wales*
*E-mail:* `lewisf@cf.ac.uk`

**David M. Russell, Rob P. Fender**
*School of Physics and Astronomy,*
*University of Southampton, Highfield, Southampton, England.*
*E-mail:* `davidr@phys.soton.ac.uk, rpf@phys.soton.ac.uk`

**Paul Roche**
*Faulkes Telescope Project*
*School of Physics and Astronomy, Cardiff University, 5 The Parade, Cardiff, Wales*
*E-mail:* `paul.roche@faulkes-telescope.com`

The Faulkes Telescope Project is the educational arm of the Las Cumbres Observatory Global Telescope Network (LCOGT). It currently has two 2-metre robotic telescopes, located at Haleakala on Maui (FT North) and Siding Spring in Australia (FT South). It is planned to increase this to six 2-metre telescopes in the future, complemented by a network of 30-40 smaller (0.4 – 1 metre) telescopes providing 24 hour coverage of both northern and southern hemispheres.

We are undertaking a monitoring project of 10 low-mass X-ray binaries (LMXBs) using FT North to study the optical continuum behaviour of X-ray transients in quiescence. The introduction of FT South in September 2006 allows us to extend this monitoring to include 17 southern hemisphere LMXBs. With new instrumentation, we also intend to expand this monitoring to include both infrared wavelengths and spectroscopy.

# 1. Introduction

Las Cumbres Observatory, Inc. is a non-profit private foundation that is building two networks of telescopes, scientific and educational.

There will be a network of six to eight 2-metre telescopes primarily for scientific research spaced around the world with the use of high quality instruments. Because of their planned longitudinal spacing, observations will be able to occur all day, everyday and can focus on time-variable astronomy observations.

The educational network will contain about 30 telescopes ranging in size from 0.4 to 1.0 metre. These telescopes will also be spaced around the world with high quality instruments.

The aim of the LCOGT is to provide two complete, longitudinally-distributed rings of research telescopes, one in the northern hemisphere, and one in the south. This will allow us to follow the time variations of single objects for days or weeks.

Wherever and whenever a short-lived event occurs, LCOGT will have at least one telescope that is in the dark and able to observe it. It is intended to operate the LCOGT network as a single distributed instrument, with the telescopes being run robotically and with the observing schedule chosen to be responsive both to overall scientific goals and to targets of opportunity.

Many aspects of astronomy are natural targets for a global network, ranging from asteroids, extrasolar planets and X-ray binaries to supernovae and gamma-ray bursts.

# 2. Project Aims

The monitoring campaign started early in 2006 and has two primary aims.

## 2.1 To identify transient outbursts in LMXBs

LMXBs may brighten in the optical/near-infrared for up to a month before X-ray detection. The behaviour of the optical rise is poorly understood, especially for black hole X-ray binaries. Catching outbursts from quiescence will allow us to examine this behaviour and alert the astronomical community to initiate multi-wavelength follow-up observations.

## 2.2 To study the variability in quiescence

Recent results have suggested that many processes may contribute to the quiescent optical emission, including emission from the jets in black hole systems (e.g. Russell et al. 2006) [1]. By monitoring the long-term variability of quiescent LMXBs, we will be able to provide constraints on the emission processes and the mass functions.

## 3. Preliminary Results from A 0620-00

The FT North light curve of A 0620-00 indicates waveband-correlated variability (Fig. 1). By folding the data on the known orbital period of the binary (7.7523 hours) [2], clear periodic behaviour is apparent of ~0.3 mag in amplitude (Fig. 2).

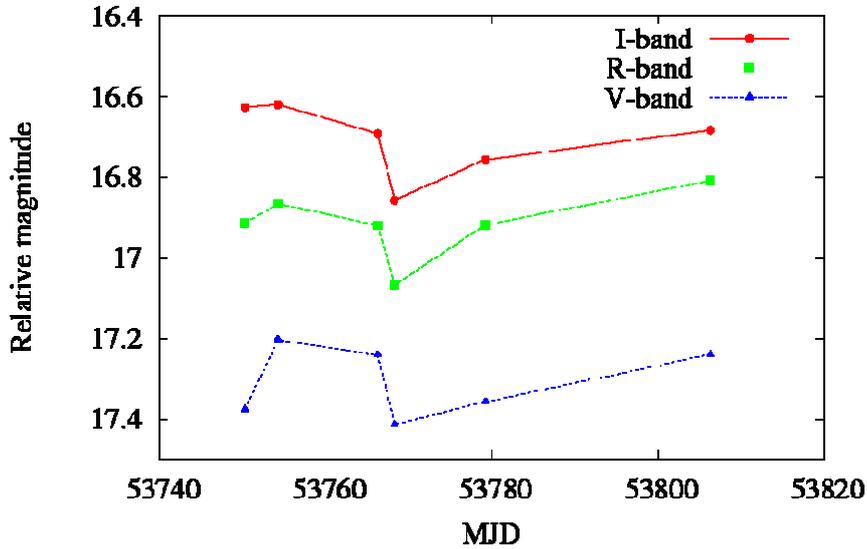

**Figure 1** Lightcurve for A 0620-00

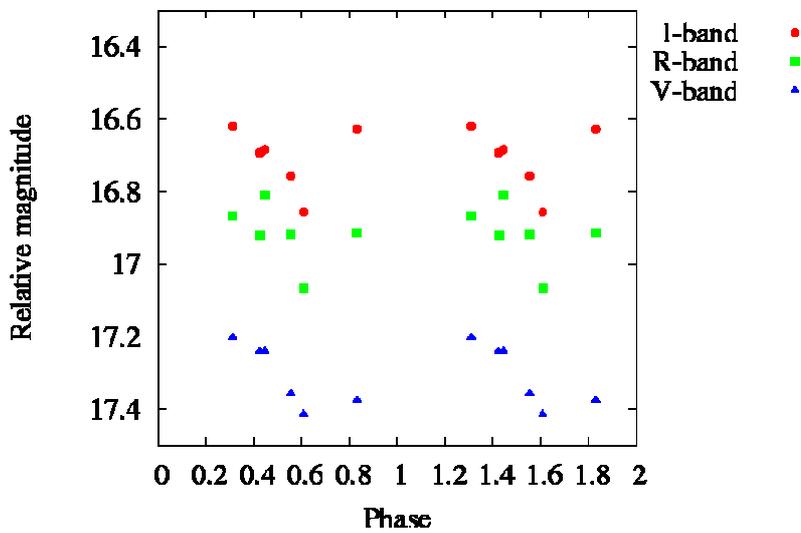

**Figure 2** Lightcurve for A 0620-00 folded on the known orbital period

## 4.     Preliminary Results from 4U 1957+11

Variability is evident in the light curve of 4U 1957+11 (Fig. 3); its behaviour is similar in all three bands. This black hole candidate X-ray binary has complex periodic variability on the known 9.33 hour orbital period [3]. The colour-magnitude diagram of this source shows a redder spectrum at higher luminosity, with one inconsistent data point (Fig. 4). This behaviour is not expected for emission from the accretion disc, which is expected to be bluer (and hotter) at high luminosities. More than one emission process could be responsible, for example, the optically thin part of the jet spectrum (which is red); clues will come from the continued monitoring.

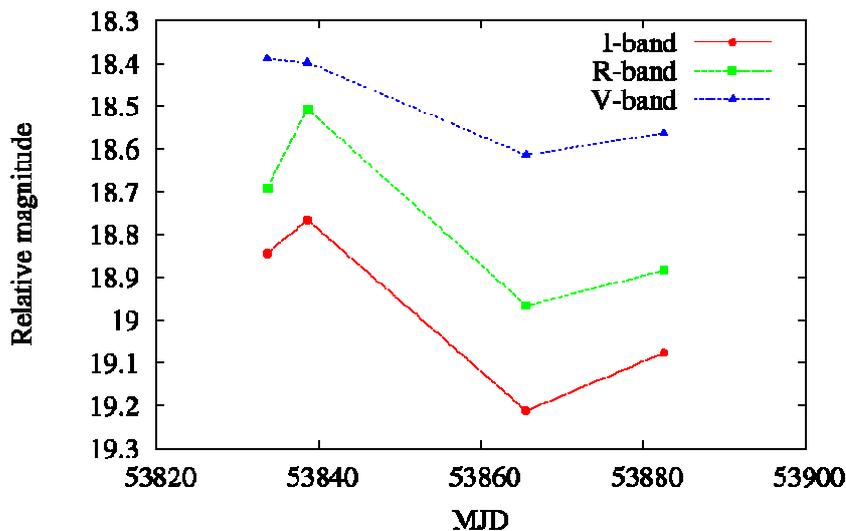

**Figure 3** Lightcurve for 4U 1957+11

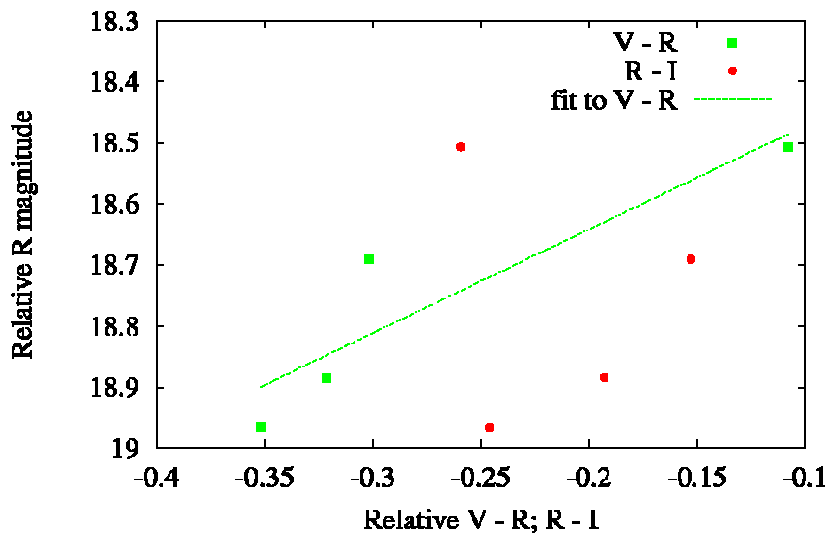

**Figure 4** Colour-magnitude diagram for 4U 1957+11

## 5. Target Lists

X-ray transients (mostly black hole systems) were selected to be targets; we focused on the sources with luminous optical counterparts which are most likely to go into outburst.

### 5.1 FT South Targets

| | |
|---|---|
| XTE J0929-314 | GRS 1009-45 |
| GRS 1124-68 | GS 1354-64 |
| Cen X-4 | 4U 1543-47 |
| XTE J1550-564 | 4U 1608-52 |
| XTE J1650-500 | GRO J1655-40 |
| GX 339-4 | H 1705-250 |
| GRO J1719-24 | XTE J17464-3213 |
| SAX J1808.4-3658 | XTE J1814-338 |
| HETE J1900.1-2455 | |

### 5.2 FT North Targets

| | |
|---|---|
| GRO J0422+32 | 4U 0614+09 |
| A 0620-00 | XTE J1118+480 |
| XTE J1859+226 | Aql X-1 |
| 4U 1957+11 | GS 2000+25 |
| V404 Cyg | XTE J2123-058 |